\documentclass[acmtog,screen, authorversion]{acmart}

\AtBeginDocument{%
  \providecommand\BibTeX{{%
    \normalfont B\kern-0.5em{\scshape i\kern-0.25em b}\kern-0.8em\TeX}}}

\setcopyright{acmlicensed}
\copyrightyear{2024}
\acmYear{2024}
\acmDOI{XXXXXXX.XXXXXXX}

\acmConference[Conference acronym 'XX]{Make sure to enter the correct
  conference title from your rights confirmation email}{June 03--05,
  2018}{Woodstock, NY}
\acmBooktitle{Woodstock '18: ACM Symposium on Neural Gaze Detection,
 June 03--05, 2018, Woodstock, NY} 
\acmISBN{978-1-4503-XXXX-X/18/06}

\usepackage[whole]{bxcjkjatype}
\usepackage[unicode]{hyperref}

\usepackage{xcolor}
\usepackage{graphicx}

\usepackage{amsmath}
\usepackage{breqn}
\usepackage{caption}
\usepackage{comment}
\usepackage{docmute}
\usepackage{empheq}
\usepackage{enumitem}
\usepackage{here}
\usepackage{karnaugh-map}
\usepackage{lscape}
\usepackage{multirow}
\usepackage{setspace}
\usepackage{siunitx}
\usepackage{url}

\usepackage{twemojis}
\usepackage{array}
\usepackage{amssymb} 
\usepackage{pifont}  
\usepackage{xcolor}  
\usepackage{algorithm}
\usepackage{algorithmic}
\usepackage{booktabs}
\usepackage{subcaption}
\usepackage[english]{babel}  
\usepackage{graphicx}  

\definecolor{ForestGreen}{RGB}{34,139,34}

\begin{document}

\title{AnimeGaze: Real-Time Mutual Gaze Synthesis for Anime-Style Avatars in Physical Environments via Behind-Display Camera}

\author{Kazuya Izumi}
\email{izumin@digitalnature.slis.tsukuba.ac.jp}
\affiliation{%
  \institution{University of Tsukuba}
  \country{Japan}
}

\author{Shuhey Koyama}
\email{Shuhey@digitalnature.slis.tsukuba.ac.jp}
\affiliation{%
  \institution{Digital Nature Group}
  \country{Japan}
}

\author{Yoichi Ochiai}
\email{wizard@slis.tsukuba.ac.jp}
\affiliation{%
  \institution{\mbox{R\&D Center for Digital Nature}}
  \country{Japan}
}


\begin{abstract}
Avatars on displays lack the ability to engage with the physical environment through gaze. To address this limitation, we propose a gaze synthesis method that enables animated avatars to establish gaze communication with the physical environment using a camera-behind-the-display system. The system uses a display that rapidly alternates between visible and transparent states. During the transparent state, a camera positioned behind the display captures the physical environment. This configuration physically aligns the position of the avatar's eyes with the camera, enabling two-way gaze communication with people and objects in the physical environment. Building on this system, we developed a framework for mutual gaze communication between avatars and people. The framework detects the user's gaze and dynamically synthesizes the avatar's gaze towards people or objects in the environment. This capability was integrated into an AI agent system to generate real-time, context-aware gaze behaviors during conversations, enabling more seamless and natural interactions. To evaluate the system, we conducted a user study to assess its effectiveness in supporting physical gaze awareness and generating human-like gaze behaviors. The results show that the behind-display approach significantly enhances the user's perception of being observed and attended to by the avatar. By bridging the gap between virtual avatars and the physical environment through enhanced gaze interactions, our system offers a promising avenue for more immersive and human-like AI-mediated communication in everyday environments.
\end{abstract}



\begin{teaserfigure}
  \includegraphics[width=\textwidth]{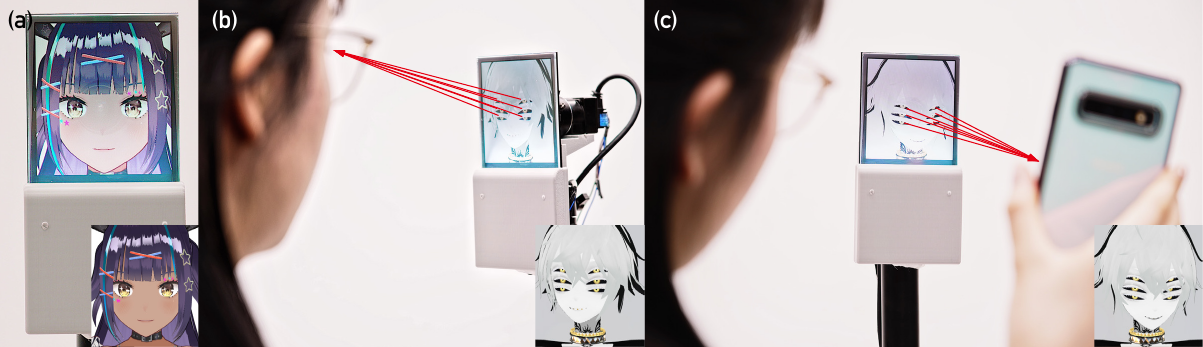}
    \caption{An avatar on the display can gaze at a location in the physical space. (a) An avatar with two eyes looking straight ahead. (b) An avatar with six eyeballs looking at the user's face. (c) An avatar with six eyeballs looking at the user's smartphone.}
  \Description{teaser}
  \label{fig:teaser}
\end{teaserfigure}


\maketitle

\section{Introduction}

Although the rise of large language models has made AI verbal communication more popular, AI still cannot gaze at us.

Within the graphics community, gaze synthesis and representation remain critical research areas, particularly for applications involving animated human face models~\cite{Lee2002-rx}, eye contact in agents~\cite{Kipp2008-oa}, and joint attention techniques~\cite{Courgeon2014-yf, Jording2018-sh}.

Recent advancements have introduced methods for synthesizing human-like eye movements in conversational AI agents \cite{Canales2023-il, Dembinsky2024-my, Dembinsky2024-zd} and generating realistic animations for in-game characters \cite{Jin2019-rx, Pan2020-kd, Pan2024-la}. These efforts aim to overcome the ``uncanny valley'' by creating more natural gaze cues. Additionally, the development of non-human avatars has expanded the possibilities of gaze interaction, introducing designs with multiple or unconventional eyes. This shift broadens the discussion of gaze synthesis to include more diverse forms of eye-based communication.
Despite this progress, existing research largely focuses on virtual environments and overlooks gaze interactions involving real-world objects. AI agents displayed on screens can now interpret real-world visual information for language communication, yet they struggle to detect or respond to a user's gaze directed at them. This limitation poses a challenge for achieving mutual gaze communication between on-screen avatars and users sharing the same physical space.

In contrast, human–human telepresence systems have extensively studied mutual gaze alignment using behind-display cameras that align the user's line of sight with the camera's optical axis \cite{Ishii1992-ib, Okada1994-oa, Otsuka2016-wx}. Izumi et al. \cite{Izumi2024-gg} demonstrated that such ``eye-contact displays'' could enhance human–AI interactions, though achieving more advanced gaze behaviors (e.g., joint attention) remains a challenge.

This paper addresses these gaps by leveraging a behind-display camera as part of a hardware platform that enables mutual gaze communication between users and avatars. This setup physically aligns the avatar's eyes with the camera's position, allowing the avatar to detect where the user is looking in real-world space and respond with accurate gaze cues.

By aligning the user's view with the avatar's perspective, as described by Izumi et al. \cite{Izumi2024-gg}, we achieve accurate gaze exchange without requiring computationally intensive processing. To further improve this, we incorporate a wide-angle camera, enabling avatars to interact not only with a single user but also with nearby individuals or objects.

We also introduce a novel calibration procedure to enhance gaze accuracy. The avatar perceives its 3D surroundings by mapping 2D camera images to real-world coordinates, using lens parameters to perform this transformation. Since avatars' gaze targets are rendered on a 2D display, discrepancies between the intended and perceived gaze direction can arise. To address this, developers calibrate the system by recording differences between the avatar's nominal gaze point and the user's subjective sense of eye contact. This process ensures that the avatar's gaze consistently aligns with real-world objects or individuals.

The contributions of this work include:

\begin{enumerate}
    \item A method for achieving mutual gaze between humans and AI avatars, leveraging a transparent display and behind-display camera.
    \item A formalized approach to synthesizing diverse gaze behaviors, including those for non-human or multi-eyed avatars.
    \item A calibration technique to minimize the ``Mona Lisa effect'' for flat-panel avatar displays.
\end{enumerate}
\section{Problem Statement}

\begin{figure}[htbp]
    \centering
    \includegraphics[width=\linewidth]{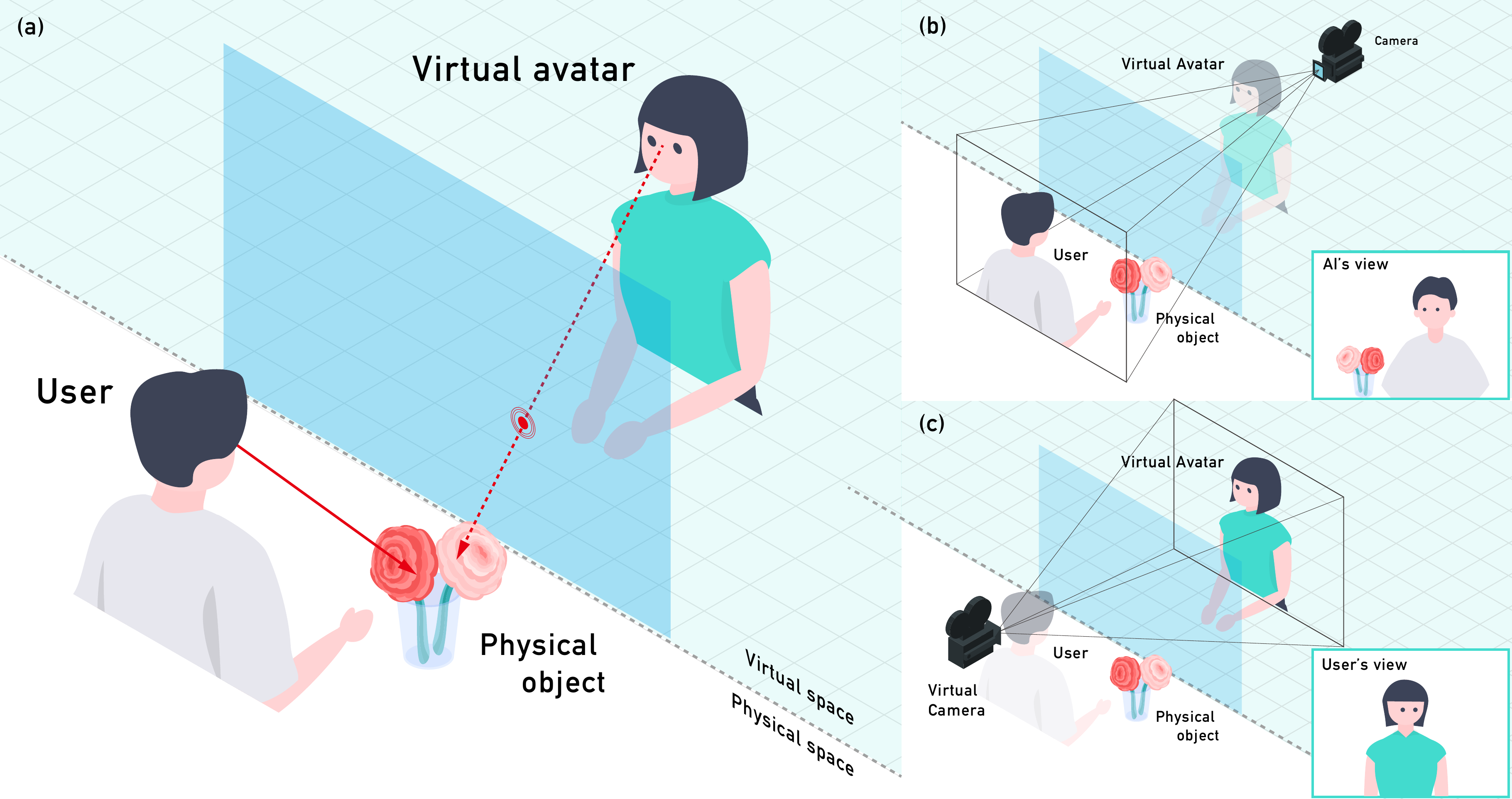}
    \caption{The Gaze Interaction Space in Our Problem Statement. (a) When a virtual avatar gazes at a physical object over the display, (b) the avatar perceives the physical space as a two-dimensional plane from the camera, and (c) the user perceives the virtual space as a two-dimensional plane from the virtual camera.}
    \label{fig:problem_statement}
\end{figure}

This paper examines gaze transmission between humans and AI avatars using a display and monocular camera setup akin to a kiosk system, as illustrated in \autoref{fig:problem_statement}. Information from the physical and virtual spaces is compressed into two-dimensional planes via their respective cameras and presented to the other party. The aim and contribution of this section are to formulate the problem of gaze recognition in this context and to extend conventional gaze and eye movement designs to avatars with diverse eye configurations.

\subsection{Fixation by Avatars with Various Eye}

In this context, avatars are considered 3D character models equipped with eyes. These avatars may be human or non-human, and the number of eyeballs is not necessarily two. This section formulates whether a user can recognize the fixation point when an avatar with several eyeballs different from the two engages in gaze communication by fixating on a point.

Generally, gaze synthesis in avatars is defined by the rotational movement of the eyeballs [Reference]. In the task of fixating on an object, once the fixation vector \( \mathbf{v} \) and the distance \( d \) from the fixation point are specified, the rotation of the eyeballs is uniquely determined. Therefore, the gaze synthesis task for fixation behavior can be replaced by the task of determining \( \mathbf{v} \) and \( d \).

Thus, when an avatar with \( N \) eyeballs fixates on a point, whether the user can recognize the fixation point can be formulated as follows.

\subsection{Avatar's Fixation Through the Screen}

In the setup presented in this paper, both the user and the avatar conduct gaze communication through planes that capture the counterpart's space in two dimensions. When the avatar fixates on an object located at coordinates \( (u, v) \) in the camera image, and if the camera matrix and distortion coefficients are known, the avatar needs to recognize the user's gaze and direct its own gaze by computing the vector of a ray cast from the origin of the camera coordinates to \( (u, v) \), along with the avatar's gaze vector.

Consider a monocular camera with internal parameter matrix \( \mathbf{K} \) (with distortion coefficients either known or pre-corrected). When pixel coordinates \( \mathbf{x} \) are given on the image captured by this camera, it is assumed that an unknown object is depicted at that location. Furthermore, in the scene, there are 3D points \( \mathbf{X} \) with known positions in the world coordinate system, each corresponding to 2D image points \( \mathbf{x} \) (the basic condition of the Perspective-n-Point (PnP) problem).

In this study, in addition to estimating the external parameters \( R \) and \( \mathbf{t} \) (rotation and translation vectors) of the monocular camera, the goal is to estimate the 3D point \( \mathbf{X}_{\mathrm{obj}} \) corresponding to the object's coordinates on the screen. Furthermore, by setting the avatar's eyeball center as \( \mathbf{A} \) in the world coordinate system and combining constraints that enable the avatar to naturally look at this object, the problem considers simultaneously optimizing \( R \), \( \mathbf{t} \), and \( \mathbf{X}_{\mathrm{obj}} \).

Specifically, parameters that simultaneously satisfy the following three requirements are sought:

\begin{itemize}
    \item The known 3D points \( \mathbf{X} \) and their corresponding image points \( \mathbf{x} \) align such that the PnP error is minimized by the camera's external parameters \( R \) and \( \mathbf{t} \).
    \item When the unknown object \( \mathbf{X}_{\mathrm{obj}} \) depicted at \( (u, v) \) on the screen is projected using the above external parameters, the reprojection error is minimized.
    \item The avatar is looking at the object; that is, \( \mathbf{X}_{\mathrm{obj}} \) is optimized to be close to a certain ideal gaze direction \( \mathbf{v}_A \) (or adheres to a specific gaze control policy).
\end{itemize}

y integrating these requirements, the problem can be formulated as the minimization of the following objective function:

\[ 
f = \min_{R, \mathbf{t}, \mathbf{X}_{\mathrm{obj}}} \left( E_{\text{PnP}}(R, \mathbf{t}) + E_{\text{reproj}}(\mathbf{X}_{\mathrm{obj}}, R, \mathbf{t}) + E_{\text{gaze}}(\mathbf{X}_{\mathrm{obj}}, \mathbf{v}_A) \right)
\]

If the optical center of the camera \( \mathbf{C} \) and the avatar's eyeball position \( \mathbf{A} \) are at the same coordinates, it can be assumed that \( \mathbf{C} = \mathbf{A} \). In this case, the ray used when reprojecting \( \mathbf{X}_{\mathrm{obj}} \) (the straight line from the camera center to the object) and the avatar's gaze (the straight line from the eyeball center to the object) are the same line segment. That is,

\[ \mathbf{v}_{A}(\mathbf{X}_{\mathrm{obj}}) = \frac{\mathbf{X}_{\mathrm{obj}} - \mathbf{C}}{\|\mathbf{X}_{\mathrm{obj}} - \mathbf{C}\|} \]

Thus, the direction vector in the camera coordinate system and the avatar's gaze vector completely coincide, causing the term related to \( \mathbf{v}_{A}(\mathbf{X}_{\mathrm{obj}}) \) in the objective function to naturally negate its angular component (or the influence of reprojection).

As a result, the estimation of the object's depth \( \mathbf{X}_{\mathrm{obj}} \) is determined almost solely by depth estimation via the PnP problem (or ray casting), eliminating the need to consider discrepancies in gaze direction (since the gaze and camera ray coincide), which significantly simplifies the algorithm.

In practice, although the depth cannot be precisely determined from monocular camera images, if the camera origin and the avatar's eyeball coincide, even an imprecise estimation of \( \mathbf{X}_{\mathrm{obj}} \) incurs minimal discomfort regarding gaze direction. Users can intuitively recognize that the avatar is "looking at an object on the screen," and since they are not particularly conscious of the depth dimension, practical scenarios for creating a sense of \textbf{joint attention} encounter fewer issues.

\subsection{Addressing the Mona Lisa Effect}

\begin{figure}[htbp]
    \centering
    \includegraphics[width=\linewidth]{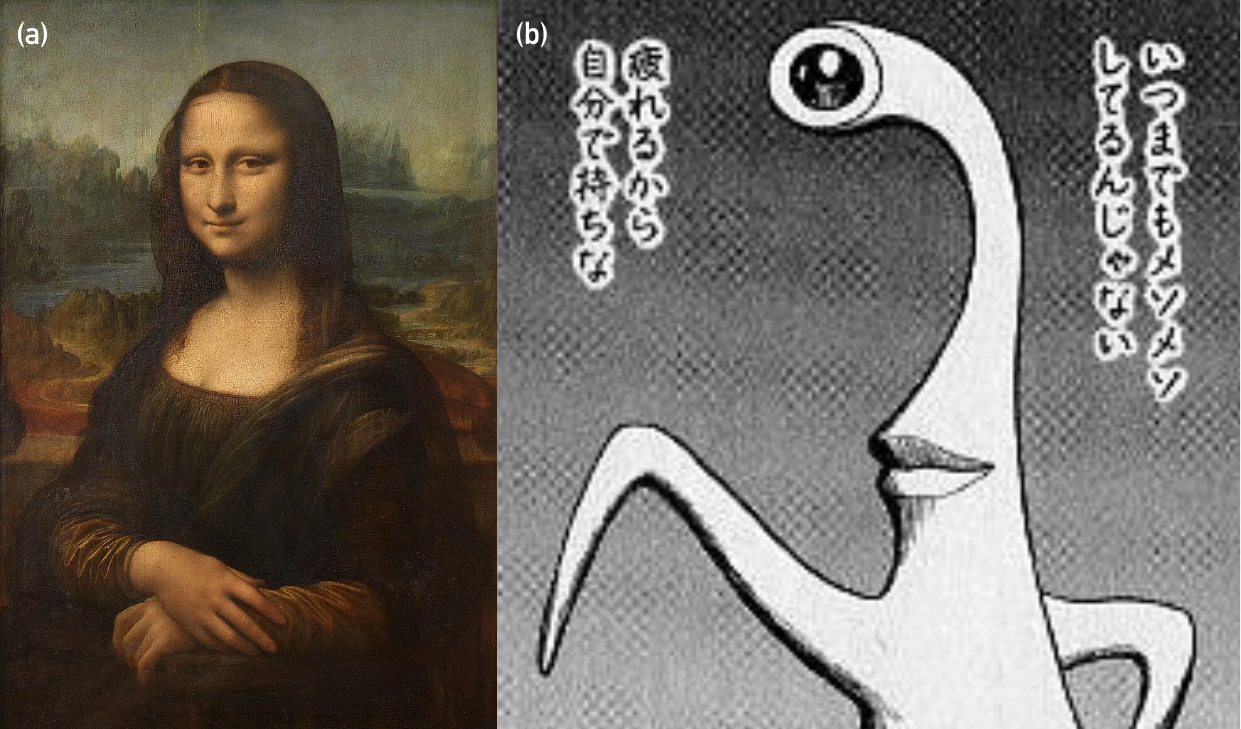}
    \caption{(a) The Mona Lisa Effect in the era of the Mona Lisa, (b) The diversification of the Mona Lisa Effect in the modern era (quoted from Kiseiju, Volume 10).}
    \label{fig:monalisa}
\end{figure}

Generally, faces displayed on a flat plane induce an illusion known as the Mona Lisa effect, where the observer feels that the face is looking straight at them from any angle, as depicted in \autoref{fig:monalisa}~\cite{Mitake_undated-ew, Moubayed2012-ll}. It is known that the same illusion occurs even when the avatar displayed on a flat plane is facing forward~\cite{Kum2024-pg}, and in the setup of this paper, this is a problem that requires attention.

Even if the avatar directs its gaze three-dimensionally in virtual space, when rendered on a flat display, users may perceive the gaze as a different vector. Furthermore, this discrepancy in gaze perception varies depending on the type of avatar. Therefore, a method for correcting the gaze vector for each avatar to align with the user's perception is discussed.

In this context, the correction task is reframed as minimizing the discrepancy between the avatar's gaze vector and the gaze vector perceived by the user. These algorithms can be formulated as follows:

\[ 
\text{Minimize} \quad E_{\text{perception}} = \| \mathbf{v}_{A}^{\text{perceived}} - \mathbf{v}_{A}^{\text{actual}} \|
\]

This study employs an original calibration method, described later, to minimize the user's misperception of the avatar's gaze. By performing this minimization to appropriately adjust the avatar's gaze on the screen and enabling interactive gaze synthesis, the goal is to reduce the Mona Lisa effect using only a flat display setup.

\section{Related Work}

\begin{figure}[htbp]
    \centering
    \includegraphics[width=\linewidth]{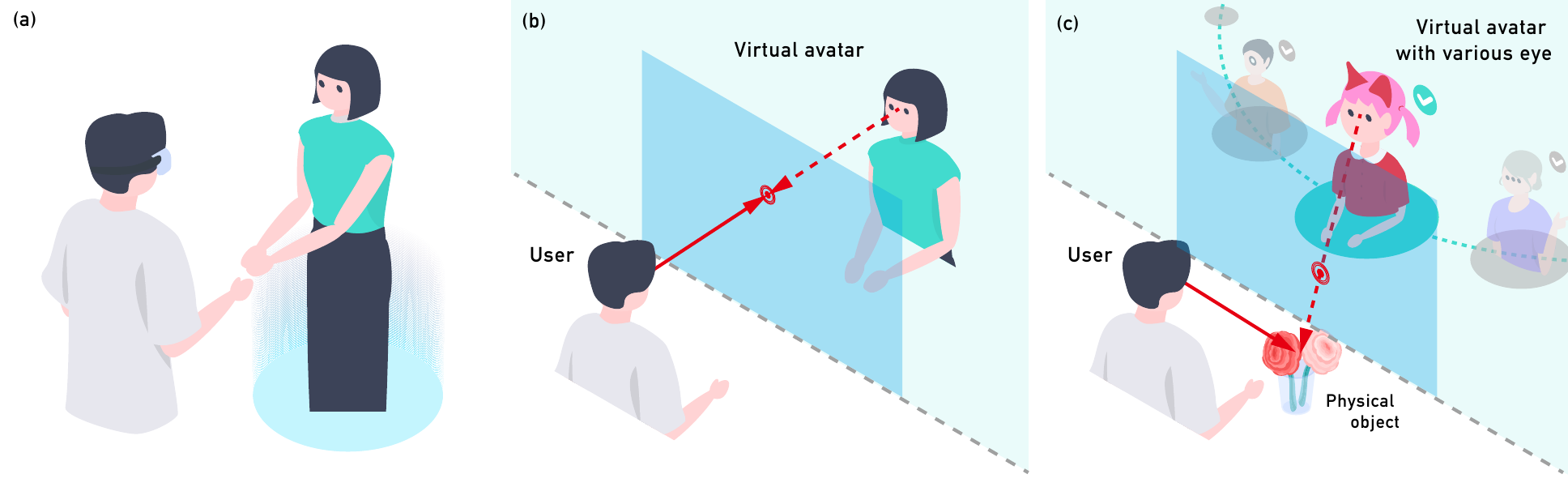}
    \caption{Position of this paper. (a) The user can make eye contact with a virtual avatar in VR space. (b) The user can make eye contact with the virtual avatar through the screen. (c) This paper is a system in which avatars with various eye characteristics can communicate with each other in physical space.}
    \label{fig:related_works}
\end{figure}

In this section, we will organize eye contact communication with AI avatars, which has been conducted using human models.
Until now, eye contact with avatars has mainly been discussed in full virtual spaces such as VR spaces or when the avatar faces forward unilaterally, as shown in \autoref{fig:related_works}(a, b).
As shown in \autoref{fig:related_works}(c), the contribution of this paper is to propose a method for physically correct eye contact communication between avatars and users, and to study eye design methods that are independent of the number and shape of the avatar's eyeballs.

\subsection{Gaze Cues in Human-Agent Interaction}
In the field of Human–Agent Interaction (HAI), numerous methods have been proposed to treat avatars as ``faces'' of an agent for the purpose of gaze-based communication with users~\cite{Garau2001-lg}. Gaze behaviors in conversation include eye contact, joint attention, and gaze aversion. In particular, various approaches have synthesized gaze for virtual agents to enhance user engagement through eye contact~\cite{Kipp2008-oa}, leverage the user's gaze tracking for joint-attention-based interactions~\cite{Courgeon2014-yf}, or introduce natural gaze aversion to avoid the discomfort that constant staring can induce. Furthermore, more fine-grained gaze behaviors have also been investigated, such as modeling pupil constriction when an avatar experiences fear~\cite{Dong2022-fw}.

This line of research on gaze communication in HAI extends beyond flat-screen avatars to include gaze interaction with avatars in virtual reality (VR) settings~\cite{Duguleana2014-td, Rogers2022-to, Suk2023-qf, Cuello-Mejia2023-nq}. Recently, constructing large gaze datasets of conversational AI agents has attracted attention, enabling the rendering of more natural gaze patterns during dialogues~\cite{Dembinsky2024-my, Dembinsky2024-zd, Richard2020-kd}. Additionally, there have been attempts to estimate visual saliency in first-person conversational footage to generate plausible gaze patterns for avatars~\cite{Boccignone2020-kh, Pan2024-la}.

Our study seeks to build upon these works by shifting from purely virtual environments to physically connected avatar scenarios, emphasizing eye contact and joint attention. Moreover, little attention has been paid to non-human avatars, particularly those that do not have exactly two eyes. By accommodating diverse “eye” configurations and enabling avatars to gaze outside the screen, we aim to provide an initial exploration of broader gaze interaction possibilities.

\subsection{Transparent Display and Behind-Display Camera}

Positioning a camera behind a display has been investigated for various purposes, such as enabling direct interaction from the rear side in computing interfaces~\cite{Wilson2005-sc, Lindlbauer2014-il} or addressing gaze mismatch problems in communication environments~\cite{Jaklic2017-dm}.

In particular, behind-display camera systems have been extensively discussed since the pioneering work of Hiroshi et al.\cite{Ishii1992-ib, Otsuka2016-wx, Okada1994-oa, Lim2021-ci} as a way to enhance gaze awareness in remote communication. With advancements in display and camera resolution, these setups have also been proposed for AR-mirror-like applications, where the user's own camera-captured image is reflected in real time\cite{Wang2024-gw}.

Furthermore, Izumi et al.~\cite{Izumi2024-gg} demonstrated that a compact enclosure could facilitate everyday use of eye-contact displays, suggesting their potential for gaze communication not only between humans but also between humans and AI avatars. This versatility has been enabled by the improved capability to capture high-resolution images from a behind-display camera, thus making it suitable for a wide range of display content.

Our research leverages such behind-display camera systems to facilitate gaze communication between humans and AI avatars in the physical world, aiming to address the challenges discussed in Section~2 with low computational cost and a straightforward hardware configuration.
\section{Implementation}

\subsection{Physical Alignment of the Camera and Avatar's Eyeballs}

\begin{figure}[htbp]
    \centering
    \includegraphics[width=\linewidth]{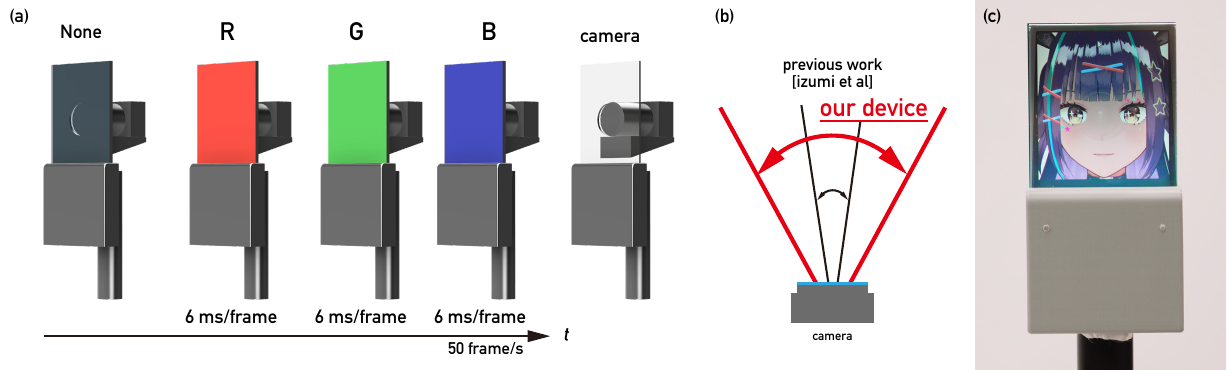}
    \caption{Hardware Configuration}
    \label{fig:hardware}
\end{figure}

As described in Section 2, directing the avatar's gaze into physical space requires estimating the depth of objects in camera images and reconstructing the spatial relationships within the virtual environment. The present research, following the framework of Izumi et al.~\cite{Izumi2024-gg}, employed an eye-contact display composed of a transparent display that alternates between transmitting and scattering light in its liquid crystal layer, with a camera positioned behind it.

To address the problem outlined in Section 2.2, an eye-contact display was utilized according to the configuration of Izumi et al. This display incorporates a transparent screen that switches between transparent and scattering states in its liquid crystal layer and a camera placed behind the display. The field-sequential drive of the device operates at 180 Hz, while the display's overall refresh rate is 50 Hz. During the transparent state, the field-sequential drive prompts the camera to capture images. The exposure time is configured at 6 ms within each frame, with a frame rate of 50 frames per second and a total exposure of 20 ms, thereby preventing interference from the display's content when capturing the user's face. The camera streams video at a resolution of 1440×1080 pixels and provides images as a virtual camera feed using DirectShow Filters. The integrated camera is the BU160MCF, produced by Toshiba Teli Corporation. The transparent display is a 4-inch full-color LCD with a resolution of 320×360 pixels, as reported in~\cite{Okuyama2017-eh, Okuyama2021-ml}.

Although the camera described in Izumi et al. had a field of view of about 30 degrees, the field of view in the current setup has been increased to \(N\) degrees, and the resolution of the captured images has notably improved. This upgrade enables the camera to capture physical environments beyond the user's face, expanding potential interactions.

\begin{figure}[htbp]
    \centering
    \includegraphics[width=\linewidth]{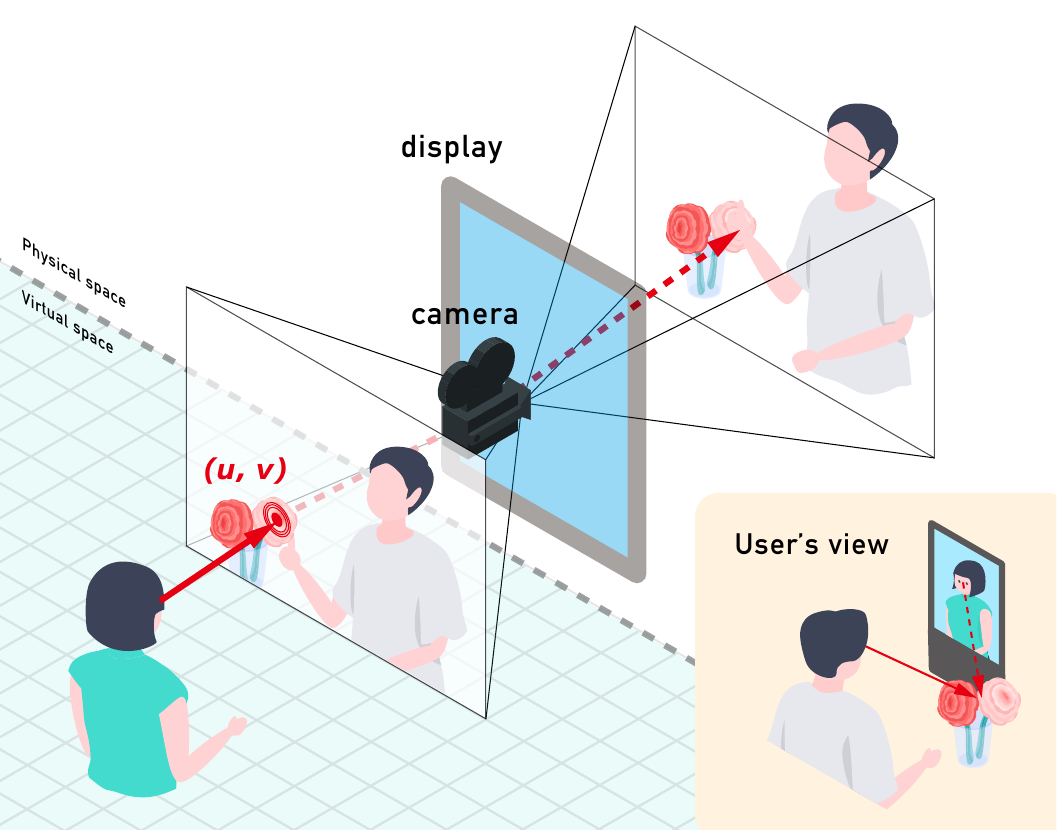}
    \caption{Gaze Interaction Space between User and Avatar. The avatar's line of sight and the camera's optical axis are aligned by the eye contact display, and the target physical object exists at one of the points of the avatar's gazing vector.}
    \label{fig:virtual_space}
\end{figure}

An AI avatar was displayed on this transparent screen to build a gaze interaction environment, as shown in \autoref{fig:virtual_space}. This environment was implemented in Three.js using millimeter-based coordinates. Initially, the positional relationship among the avatar, the virtual camera, and the image plane observed by the avatar was defined. Camera calibration was then performed with a checkerboard pattern, yielding the camera matrix \( M \) and distortion coefficients \( c \) as shown in Equations N and M.

The virtual camera and image plane were positioned at distance \( f \) from the avatar's eyeballs and oriented toward the avatar's face. The size of the image plane, \( W \times H \) mm, was calculated as follows:

\[
W = \frac{W_d}{W_o} \times W_i, \quad H = \frac{H_d}{H_o} \times H_i
\]

where \(W_i \times H_i\) pixels is the camera resolution, \(W_d \times H_d\) mm is the physical display's size, and \(W_o \times H_o\) pixels is the display's resolution. The virtual camera's parameters match those of the camera placed behind the display.

Under this configuration, if the avatar fixates on \((u, v)\) on the image plane, the avatar is rendered so that it appears to be gazing at a position beyond that plane in physical space. It is assumed that distortion correction has been applied to the image plane displayed to the avatar.

\subsection{Correction of the Avatar's Gaze Deviation}




\subsubsection{Calibration Algorithm}

\begin{figure}[htbp]
    \centering
    \includegraphics[width=\linewidth]{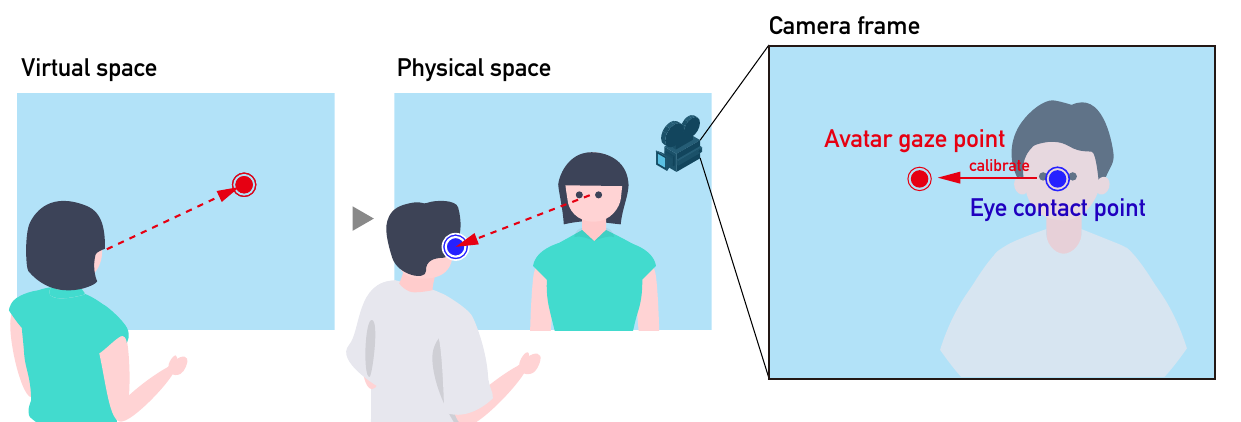}
    \caption{Calibration flow. When a virtual avatar gazes at a point on the screen, the user perceives that the avatar is looking at a different point. Calibration is performed using these pairs as calibration points.}
    \label{fig:calibration}
\end{figure}

To mitigate the observed misperception, gaze calibration was performed. Developers collected calibration points while engaging with the avatar. The avatar fixated on predefined locations on the image plane, and developers recorded points where they perceived direct eye contact. These points formed pairs of ground truth fixation locations and perceived fixation locations.

When the avatar fixated on a ground truth point, users reported that they saw the avatar looking at a different, measured point. After gathering these calibration pairs, symbolic regression was carried out based on the method described by Hassoumi et al., which adjusted the avatar's gaze direction.

This algorithm ensures that when the avatar's gaze is directed into physical space, observers accurately recognize its point of fixation. Figure N(a) presents the collected calibration points, and Figure N(b) shows the validation outcomes after calibration. The method successfully corrected the gaze deviation, corroborating the preliminary study's finding that the avatar's gaze was often perceived as deviating significantly in the horizontal direction.

\section{Results}

\begin{figure*}
    \centering
    \includegraphics[width=\linewidth]{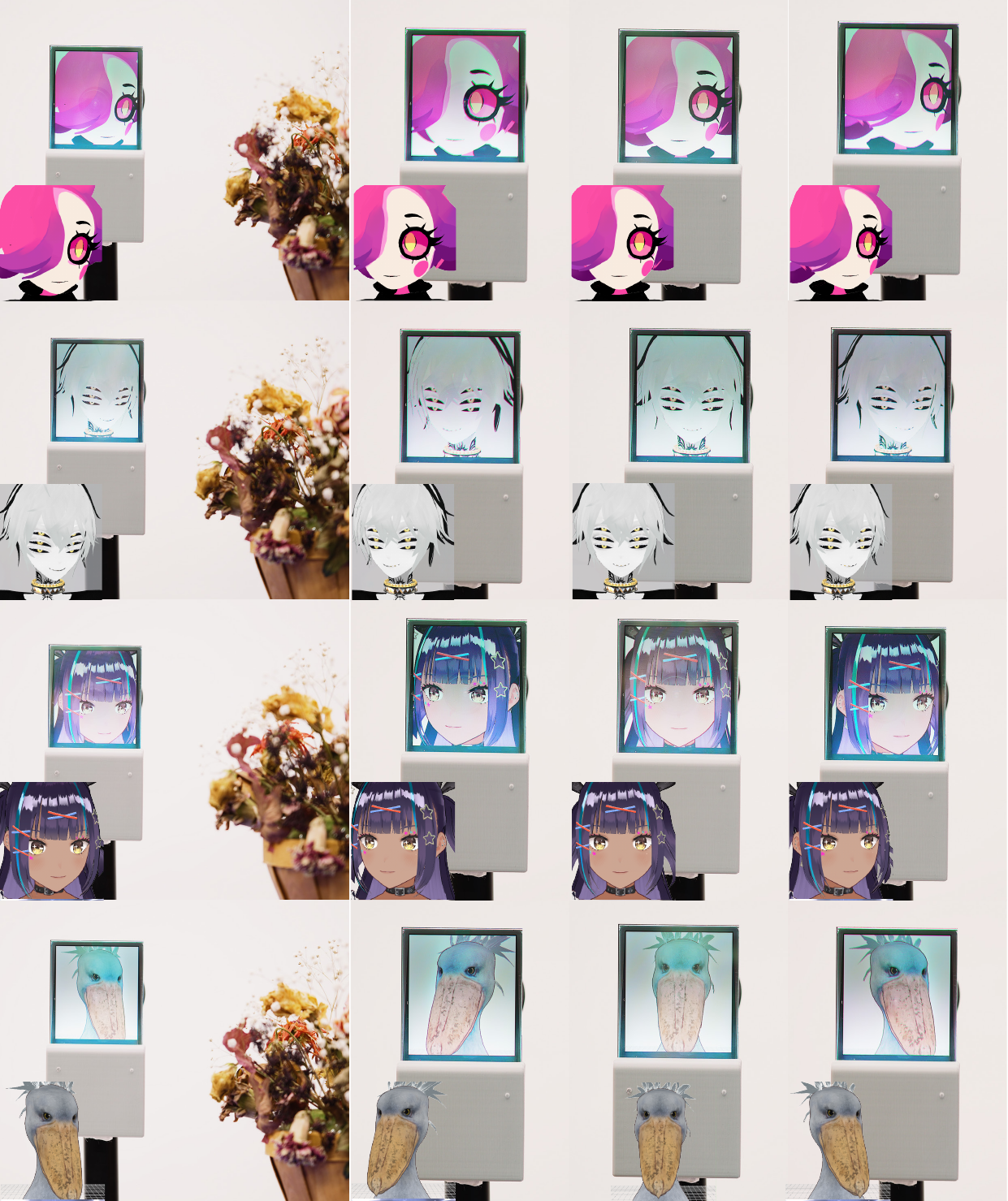}
    \caption{Results of avatar gaze recognition by our system. The avatar gazes at the right flower, the left side, the front, and the right side, respectively. Regardless of the number of eyes on the avatar, and regardless of whether the avatar is human or not, our system is able to achieve a high degree of gaze transmission.}
    \label{fig:result}
\end{figure*}

The \autoref{fig:result} shows the result of displaying an avatar on our system and synthesizing the gaze.

\section{Discussion}

\subsection{Gaze design in independent eye movements}

The present study considered only gaze design when the avatar's eyeballs were all focused on a single point. However, changes in eye contact recognition in avatars with three or more eyeballs, due to the independent movement of multiple eyeballs, were not examined. In Section 2.1, it was formulated that the user's eye contact recognition shifts according to the sum of weights reflecting each eyeball's contribution to eye contact recognition. Consequently, if the majority of the eyes are directed toward the target, gaze recognition may be perceived, and the head direction may act as a significant contributor. The definition of gaze in such non-human avatars must be reassessed, based on how each eyeball rotates.

\subsection{Hardware constraints on interaction}

In the present research, a gaze design system was constructed using a display and a monocular camera arrangement, similar to a kiosk system. The monocular camera cannot estimate the depth direction of an object, which limits the reproduction of eye gaze. However, since animated avatars do not readily perceive the depth of an object's gaze, this implementation is unlikely to introduce significant practical concerns.

Placing the camera behind the display, rather than using a depth camera, is more cost-efficient and aligns the fields of view for both the camera and the avatar. This alignment facilitates recognition of the occlusion of objects observed from the avatar's perspective. In particular, it allows the avatar to perform vision-based communication by processing the camera image with an image-to-text model, such as Vision Transformers, without requiring specialized operations to describe objects within the avatar's field of view.

The display employed in this study measures approximately four inches, which is only large enough to show the avatar's face. Nonetheless, a larger display could be developed to enable a broader range of interactions.

\subsection{Synthesizing more natural eye movement}

The principal contribution of this study is the proposal of a positioning method and a mutual gaze recognition approach to achieve eye interaction between the avatar and its physical space on the display. Discussions about how to realize gaze behavior surpassing the uncanny valley remain crucial for genuine interaction with users. The present study did not implement rhythmic talking head synthesis during speech~\cite{Canales2023-il}, synthesis of eye gaze and head movements based on one's own conversational turns~\cite{Dembinsky2024-my, Dembinsky2024-zd}, or gaze aversion after a certain period of time~\cite{Pan2024-la}, as described in previous works.

Future work will include synthesizing such natural eye movements, although it is necessary to exercise caution when extending the system to diverse avatars, including non-human avatars.

\section{Conclusion}

We proposed AnimeGaze, an eye-contact display that combines a transparent display and a behind-display camera, and a framework that enables advanced eye-contact communication that seeps into the physical space of any avatar using the display.
AnimeGaze is expected to extend the avatar gaze composition problem in the graphics community to non-human avatars and to take into account the number of eyeballs, and to enable eye contact and joint attention in physical space between the user and a planar display. The problem is expected to be extended to consider non-human avatars and the number of eyes in the community.

\begin{acks}
We are grateful to Japan Display Inc. for lending us the prototype of See-Through Face Display and Kazuhiko Sako, Kazunari Tomizawa, and Kentaro Okuyama for their technical assistance in the hardware development.
\end{acks}

\bibliographystyle{ACM-Reference-Format}
\bibliography{paperpile}

\end{document}